\documentstyle[12pt,aasms,psfig]{article}

\begin{document}
\baselineskip=12pt

\title{A Topological Extension of General Relativity}

\author{Marco Spaans}

\affil{Physics \& Astronomy Department, Johns Hopkins University}
\hfill\break
3400 North Charles Street, Baltimore, MD 21218, USA\hfill\break
Email: spaans@pha.jhu.edu

\begin{abstract}

A set of algebraic equations for the topological properties of
space-time is derived, and used to extend general relativity into the 
Planck domain. A unique basis set of three-dimensional prime manifolds is
constructed which consists of $S^3$, $S^1\times S^2$, and $T^3$. The action of
a loop algebra on these prime manifolds yields topological invariants which
constrain the dynamics of the four-dimensional space-time manifold.
An extended formulation of Mach's principle and Einstein's equivalence
of inertial and gravitational mass is proposed which leads to the correct
classical limit of the theory.

It is found that the
vacuum possesses four topological degrees of freedom corresponding to a
lattice of three-tori. This structure for the quantum foam
naturally leads to gauge groups O(n) and SU(n) for the fields, a boundary
condition for the universe, and an initial state characterized by local
thermal equilibrium. The current observational
estimate of the cosmological constant is reproduced without fine-tuning and
found to be proportional to the number of macroscopic black holes. The black
hole entropy follows immediately from the theory and the quantum corrections
to its Schwarzschild horizon are computed.

\hfill\break\noindent
{\it Subject headings:} general relativity --- quantum cosmology\hfill\break

\end{abstract}

\section{Introduction}

In the past decades much effort has been devoted to the incorporation of 
gravitation into the 
framework of quantum field theory which describes the strong and electroweak
interactions. Most work has concentrated on a formulation in terms of the
geometrical and topological properties of these interactions. Superstring
theories attempt to construct a Grand Unified Theory (GUT) in which gravity
emerges naturally as
a spin-2 excitation of the fundamental string[1]. Perturbative and
non-perturbative gravitational quantum field theory takes general relativity
as a starting point. These theories try to include quantum effects either
perturbatively or through a different formulation of Einstein gravity in
which an object more fundamental than the metric can be identified[2].
Also, an accurate description of the early
universe and the properties of black holes should derive from a quantum
theory of gravitation. Many conceptual difficulties in the unification
program and quantum cosmology have long been 
recognized: renormalizability, factor ordering, the arrow of time, and the
choice of boundary conditions[2,3,4].

It is well-known that the Einstein equations describe the geometry of
space-time, but do not
specify its topology[5]. Heuristic arguments, based on the lack of conformal
invariance in the equation of motion, suggest that on the Planck 
scale space-time should have a multiply-connected (``foam-like'') 
structure[6]. A natural
approach to incorporate the effects of a non-trivial space-time topology
is the path integral[7] over geometries, including topologically 
complicated ones[8]. To define the ground state in such a theory one should
provide for a boundary condition of the universe[9,10].

In the present work the geometrical Einstein equations are taken as a starting
point and are complemented by a theory for the topological properties of
space-time in terms of particular prime three-manifolds. The theory will be
defined on a four-dimensional topological manifold with a (3+1) split for the
dynamics. The dimension is therefore fixed.

The theory requires a choice for the three-dimensional boundary of the
universe at $t\approx 0$. In this paper the initial value, $B_i^0$, is imposed
at  $t=t_{\rm Planck}$ and no attempt is made to define a theory
at the singularity $t=0$. This work aims at constructing a theory which
constrains the topology of the universe at $t_{\rm Planck}$ and subsequently
at later times, and extends the theory of general relativity into the
Planck domain. Such an extension should not compromise the
continuous evolution of the Einstein equation on time scales much larger than 
$t_{\rm Planck}$ and should limit to a solution of the Einstein equation
for length scales much larger than $\ell_{\rm Planck}$.

This paper is organized as follows.
In Section 2 a set of algebraic equations is derived for the dynamics
of three-topologies based on a loop algebra with prime manifolds as
dynamical objects, and an 
extension of Mach's principle into the Planck domain is suggested.
Section 3 discusses the Feynman
path integral and the equivalence principle, and associates the four 
topological degrees of freedom of the vacuum with gauge groups.
In Section 4, toy models for a quantum field theory and a black hole
are presented, and the importance of topological fluctuations for the
cosmological constant and macroscopic black holes is discussed.
Finally, Section 5 contains the conclusions and a discussion of future work.
\hfill\break\noindent
The first half of the paper is rather mathematical, but all the topological
properties which are discussed, turn out to have clear physical meanings.

\section{Dynamical 3-Topologies}

\subsection{One-Loops and Two-Loops in Three-Space}

Consider the well-known superposition paradigm of quantum physics as expressed
in Feynman's path integral. Through constructive interference of wave
amplitudes along all possible paths connecting two points, a world line
is defined and associated with the trajectory of a particle.
In a topologically non-trivial manifold, the world line becomes a
plait with individual strands. When pinched at its ends, the underlying
loop structure of the plait becomes apparent. It is this structure which
reflects the topological interactions. Just like fields interact in particle
dynamics, loops interact in topological dynamics.

Planck scale phenomenology precludes the dynamical significance of other
homotopy groups in a theory for the dynamics of three-topologies.
Any one-loop in a three-manifold generates a
two-surface with the topology of a cylinder in four-dimensional space-time.
So time dilation preserves the one-loop structure. This conservation of loop
structure is required because any embedded lower-dimensional surface has a
quantum-mechanical ``width'' of the order of a Planck time.
Conversely, for the sphere $S^2$, the resulting topology is that of a cube.
So the presence of a hole in a three-dimensional cube $I^3$ inhibits the 
contraction of a two-loop, but $S^2\times I$ has a cube topology and is
dynamically trivial.

\subsection{Loop Algebra}

To facilitate the continous evolution of the Einstein equation on large scales
and for long times, and still include quantum effects, time is 
regarded as a discrete coordinate with a time step equal to the Planck time.
If one assumes that the topology of space-time on the Planck scale is 
non-trivial, then knowledge of its precise structure requires 
some formal measurement through matter degrees of freedom which cannot
measure beyond a Planck time or Planck scale without influencing the observed
system itself. Conversely, the effects of any change in topology,
although instantaneous in terms of topological invariants like Betti numbers
(kinematics), cannot be distinguished dynamically through interactions
within a Planck time. It is assumed that in the absence of matter degrees of 
freedom, this behavior holds as well.
This is a kind of semi-classical approximation in which the four-manifold with
Planck scale structure (the wave function) is separated into the Einstein
solution (a slowly varying amplitude) and the quantum foam (a rapidly varying
phase).

In analogy with the first 
order (in time) nature of the Schr\" odinger equation, any equation for the
discrete evolution of three-topologies is assumed to require only one initial 
value. Consequently, for any time $t_1$ the solution one
Planck time later can only depend on the topology at time $t_1$.

In the present work the following equation will be used to determine the
topology of the universe at times $t=(k+1)t_{\rm Planck}$ 
in the absence of matter:
$$\sum_j [T{\cal T}_i^{k+1},T^{\dagger}{\cal T}_j^{k+1}]/{\cal P}_j=
(TT^{\dagger}+T^{\dagger}T)B_i^k,\qquad k=0,1,2,... \eqno(1)$$
with $[T{\cal T}_1,T^{\dagger}{\cal T}_2]\equiv 
T({\cal T}_1\times T^{\dagger}{\cal T}_2)-
T^{\dagger}({\cal T}_2\times T{\cal T}_1)$, and
$B_i^0=\alpha_i^0 {\cal P}_i$ the three-boundary 
of the universe at $t=t_{\rm Planck}$. The basis set
${\cal B}\equiv \{{\cal P}_i\}=\{{{{\cal T}_i}\over{\alpha_i}}\}$ 
consists of prime manifolds ${\cal P}_i$ with multiplicities $\alpha_i$.
A three-manifold is called prime iff it is not the connected sum of two
three-manifolds none of which is diffeomorphic to the three-sphere.

The word three-manifold is assumed to imply closed (compact), connected and
oriented. In three dimensions any manifold $M$ is diffeomorphic to the (chiral)
unique, finite, and {\it linear} decomposition
$${\cal M}=\oplus_i {\cal T}_i=\oplus_i (\alpha_i {\cal P}_i). \eqno(2)$$
This connected sum is an associative and commutative operation in the
category of oriented three-manifolds and orientation preserving homeomorphisms.
The connected sum of any manifold $M$ and $S^3$ is homeomorphic to $M$
$$M\oplus S^3\sim M. \eqno(3)$$

\noindent The operators $T$ (annihilation) and $T^{\dagger}$
(creation) satisfy
$$T({\cal P}_1\times {\cal P}_2)=T({\cal P}_1)\times {\cal P}_2 
+ {\cal P}_1\times T({\cal P}_2), \eqno(4a)$$
$$T({\cal P}_1\oplus {\cal P}_2)=T({\cal P}_1) + T({\cal P}_2), \eqno(4b)$$
$$TT^{\dagger}({\cal P})=k{\cal P}, \eqno(4c)$$
$$T^{\dagger}T({\cal P})=l{\cal P}, \eqno(4d)$$
with $k,l\in Z.$
The $+$ rather than $\oplus$ on the right-hand side of Equations (4a) and (4b)
signifies the fact that a direct sum should not be taken at this stage, 
because ${\cal P}_1$ and ${\cal P}_2$ do not have the same dimension as 
$T({\cal P}_1)$ and $T({\cal P}_2)$.
The action of $T$ and $T^{\dagger}$ on a prime manifold ${\cal P}$ is
$$T^{\dagger}{\cal P}=S^1\times {\cal P} \eqno(5a)$$
$$T {\cal P}= \sum_i {\cal Q}_i, \eqno(5b)$$
with $TS^3=0$, i.e.~the trivial element.
The sum in Equation (5b) extends over all homotopically
inequivalent loops and ${\cal Q}$ denotes ${\cal P}$ with a loop
shrunk to a point. The summation reflects the fact that there is only
{\it one type of loop} in the theory. If the primes in question are 
chiral, i.e.~do not admit orientation reversing homeomorphisms, then
both right- and left-handed orientations, (${\cal P}$) and ($-{\cal P}$)
should be included in the decomposition (2).

\subsection{Derivation of the Equation of Motion}

The combination of three-dimensional prime manifolds and their topological
invariants associated with the discrete time loop algebra of the $T$ and
$T^{\dagger}$ operators renders Equation (1) unique, as follows.

A basis set of prime manifolds is adopted in Equation (1) because these objects
cannot be decomposed in two or more topologically non-trivial submanifolds.
Also, {\it any theory based on the concept of a
topological manifold in dimension $n$, ultimately needs to incorporate the
topological invariants of the primes in that dimension}. This is analogous to
the use of a differentiable manifold in the description of Einstein gravity.
In four dimensions the identities associated with the Riemann tensor impose
their own constraints on the dynamics of the theory.

\subsubsection{Left-Hand Side}

The left-hand side of Equation (1) contains the dynamics of the theory
and reflects the interaction between all the prime manifolds:
The total of variations in ${\cal T}_i$ with respect to all the
primes $\{ {\cal P}_j\}$. The commutator on the left-hand side reduces to the,
topologically trivial, identity
$[T,T^{\dagger}]{\cal T}_i={\cal T}_i$ for ${\cal T}_j=S^3$.

It is easy to show that the expression for the commutator in Equation (1)
is the only multi-linear combination of operators and primes which is
anti-symmetric under the simultaneous interchange of operators and primes, and
limits to the identity for the three-sphere.
The anti-symmetry property is necessary to assure that: 1) The net effect of
an interaction and its reverse is zero. 2) The sum over both
$i$ and $j$ in Equation (1) leads to a scalar expression which is invariant
under the interchange of any two ``vectors'' ${\cal T}_i$ and
${\cal T}_j$ for time $k+1$ as required by the decomposition (2). 3) This
scalar is anti-symmetric under the interchange of the non-commuting
creation and annihilation operators $T$ and $T^{\dagger}$. If the latter
property were absent than the physical system which would result from the
complete interchange of creation and annihilation processes would be described
by the identical equation of motion. Anti-symmetry in $T$ and
$T^{\dagger}$ yields a theory which has an arrow of time.

\subsubsection{Right-Hand Side}

The right-hand side of Equation (1) contains a scalar operator which
assures the conservation of the topological properties of the primes.
One seeks an operator $O$ linear in $T$ and $T^{\dagger}$, as these are the
only objects acting on the primes, which assigns a unique number to
every prime. Also, the topological invariant should not depend on the dynamics.
That is, the order in which the creation and annihilation operators act on the
prime through $O$ does not matter.

It is easy to show that the right-hand side of Equation (1) is the only
invariant, linear and symmetric in $T$ and $T^{\dagger}$ and unique to each
prime, which yields unity for the three-sphere. The latter requirement
implements the identity (3) which shows that in the decomposition (2),
the three-sphere has an effective multiplicity of unity.

{\it The primes are the fundamental building blocks of the topological
manifold and interactions among the primes cannot change their intrinsic
properties}. The right-hand side of Equation (2) then contains the number
representative of the ``prime quantum''. As such, it provides a topological
constraint on the evolution of the multiplicities of the primes.


\subsection{Some Relevant Properties of Prime Manifolds}

Most of the results quoted below can be found in[11].
A manifold is called chiral iff it does not allow for an orientation
reversing self-diffeomorphism. A three-manifold is non-chiral iff no prime in
its prime decomposition is chiral. The three-sphere is non-chiral, but
isometries (point identifications) of $S^3$ like SO(3) or its cover SU(2) are
chiral.

A three-manifold $M$ is called irreducible, if every embedded two-sphere
in $M$ bounds a three-ball. Clearly, an irreducible three-manifold is prime
and also has a trivial second homotopy group.
Interesting enough, the converse is also true with the exception of the
handle manifold $S^1\times S^2$, which is the only non-irreducible three-prime.
This is directly related to the fact that black holes can increase their
size and mass by accretion. That is, they have a ``throat''.

An additional class of primes are those with an infinite fundamental group 
which in addition are sufficiently large. The fundamental group of such a
sufficiently large three-manifold contains as a subgroup the fundamental
group of a Riemannian surface. The latter property means physically
that a non-contractible loop on an embedded surface is also not contractible
within the ambient three-manifold. A sufficient condition for an irreducible
manifold to be sufficiently large is that the first homology group is infinite.
These irreducible primes fall into the class of $K(\pi ,1)$ spaces
(Eilenberg-MacLane spaces), whose only non-vanishing homotopy
group is the first. No connected sum containing at least one
$K(\pi ,1)$ admits a Riemannian metric of everywhere positive scalar curvature.
Moreover, if it admits a nowhere negative scalar curvature metric then it
must be flat. That is, the manifold must be the three-torus or one of its
quotients[12].

A three-manifold is called nuclear iff it is the space-like boundary of a
Lorentz four-manifold with SL(2;C) spin structure, which is isomorphic with
SO(3,1;R). It can be shown that a three-manifold is nuclear iff the number of
nuclear primes in its prime decomposition is odd. Both $S^1\times S^2$ and
$T^3$ are nuclear unlike the primes $S^1\times R_g$, where $R_g$ is a
Riemannian surface of genus $g$.

The three-torus is the only known
perfect group or Steinberg group St(3;Z). This group is a central $Z_2$
extension of SL(3;Z). A perfect group is its own commutator subgroup, where the
commutator subgroup contains elements like $(xyx^{-1}y^{-1})$ to assure that
$[x,y]=0$. It is not likely that some of the less well studied three-primes
which are not sufficiently large change the uniqueness of this result.

\subsection{Interpretation}

The physical interpretation of Equation (1) is as follows. In the absence
of matter degrees of freedom and on the Planck scale, space acquires a
multiply-connected topology as a function of the discrete time coordinate
$t=(k+1)t_{\rm Planck}$. This foam-like appearance of space-time is a
direct consequence of the absence of conformal invariance in the Einstein 
equation[6].

Although Equation (1)
possesses an arrow of time, its direction is not specified. If one sees the
creation of the universe as ``something from nothing'' then the direction of
time as given in Equation (1) is to be preferred since the evolution of the
universe turns out not to depend strongly on the initial state (see \S 2.7).
A reversed arrow of time leads to an exponentially decaying solution in
Equation (7) below. This would also suggest that the initial quantum
fluctuations required in the ``something from nothing'' scenario cannot grow
and would have to be very fine-tuned.

Because $t_{\rm Planck} >0$, every 
three-manifold $M_k$ at discrete time $k$ generates a ``small'' 
four-manifold $M_k\times t_{\rm Planck}$. The solutions to Equation (1)
are envisaged to describe the topology of the universe as it evolves with time.
Therefore, the expansion of the universe needs to be provided for as well.

Black holes, associated with the $S^1\times S^2$ handles or wormholes,
increase their mass through merging or mass accretion.
Conversely, three-tori are irreducible and remain restricted to the
Planck scale. Their intrinsic size is also assumed equal $\ell_{\rm Planck}$.
So during the evolution of the universe the number of these primes per
Planckian volume cannot exceed unity, and complete merging on sub-Planck
scales is assumed. On scales smaller than $\ell_{\rm Planck}$, the concept of
a prime manifold should perhaps be replaced by an even more pritimive structure
like a Borel ring.

Equation (1) is purely topological (non-spatial) in nature. When one computes
that the multiplicity of, say, the handle manifold is $N$ at time $n$ then
these mini black holes are created randomly throughout the entire universe.
The assumption here is that the statistical properties of space-time on
the Planck scale are homogeneous.

\subsection{Basis Set and Formal Solutions}

Currently, no complete list of three-primes is available.
Especially the not sufficiently large primes are poorly studied. A basis
set will be chosen here which incorporates nuclearity and
commutativity.

The basis set in question consists of the non-chiral primes $S^1\times S^2$
($\alpha$), $S^3$ ($\beta$) and $T^3$ ($\gamma$). 1) The non-irreducible and
nuclear handle manifold appears as part of the large scale solution to the
Einstein equation, i.e.~the Schwarzschild solution. Since the solution
holds for any super-Planck scale, the three-prime should be included.
2) A unit element is required for the large scale vacuum limit.
3) All other primes are confined to
the Planck scale and are required to be nuclear for Lorentz invariance.
Also, the indistinguishability of group elements (individual loops) under the
loop algebra and the superposition principle will be shown to demand a perfect
group structure for the three-primes with the matter fields defined on them,
hence $T^3$ (see \S 3.1).

The above choice yields the following set of equations at time $k$
$$2\alpha_k^2+\alpha_k\beta_k+4\alpha_k\gamma_k=3\alpha_{k-1},\eqno(6a)$$
$$\beta_k^2+2\alpha_k\beta_k+4\beta_k\gamma_k=\beta_{k-1},\eqno(6b)$$
$$4\gamma_k^2+2\alpha_k\gamma_k+\beta_k\gamma_k=7\gamma_{k-1},\eqno(6c)$$
with general solution for $k>0$ in the absence of matter degrees of freedom:
$$\alpha_k=3^k{{\alpha_0}\over{\beta_0}}\beta_k,\eqno(7a)$$
$$\gamma_k=7^k{{\gamma_0}\over{\beta_0}}\beta_k.\eqno(7b)$$
Note that $\alpha_k/\beta_k$ and $\gamma_k/\beta_k$ are invariant under
changes ($\Delta\beta_k$) for $k>0$. In \S 3.2 it will be shown that these are
the quantities relevant for the computation of the Feynman propagator.
Despite the trivial appearance of the basis set and the solution, there are
some interesting consequences (see \S 3 and \S 4).

\subsection{Initial Values and Possible Histories}

The initial values $B_i^0$ determine the evolutionary path in the topological
state space and Equation (3) demands that the initial values are invariant
under the transformation
$$B_i^0\rightarrow B_i^0 \oplus S^3, \eqno(8)$$
with $S^3$ the homotopy 3-sphere. Therefore, only the integer ratios
$\alpha_0/\beta_0$ and $\gamma_0/\beta_0$ are initial value data.
For $k\ge 0$ the identity (8) is then satisfied manifestly, but the ratios
$x\equiv\alpha_0/\beta_0$ and $y\equiv\gamma_0/\beta_0$ still need to be
specified. Different evolutions are associated with different choices
of these ratios.

For $\ell\sim\ell_{\rm Planck}$ merging and coalescence suggest values for
$x$ and $y$ close to unity. Values $x=y=1$ seem natural if one considers
that quantum fluctuations in the metric scale like $\ell_{\rm Planck}/\ell$.
Still, $x=0$ is even more natural since the matter degrees of freedom will
create black holes. Also, if the number of nuclear primes in a three-manifold
is even, it does not bound a Lorentz manifold. {\it The choice of $(x,y)=(0,1)$
then correponds merely to the requirement of Lorentz, or rather SL(2;C),
invariance and the
superposition principle on a topological manifold}. In this sense one can
also argue that $x>1$ is likely to give rise to a closed universe which will
recollapse in a Planck time.

\subsection{Matter Degrees of Freedom}

If black holes are formed, the topology of space-time, and hence its evolution
is modified. If the black hole is described by a Schwarzschild metric
with the topology of a wormhole then its formation increases the multiplicity
of the handle manifold, $S^1 \times S^2$, by one. Likewise, the merging of
two black holes, or complete evaporation, has the opposite effect. In
general the number of handles
$h$, associated with the matter degrees of freedom, will satisfy
$${{dh}\over{dt}}=-R_e -R_m + R_f, \eqno(9)$$
where $R_e$, $R_m$, and $R_f$ denote the rates in s$^{-1}$ for
complete evaporation, merging, and formation,
respectively. The fate of the singularity upon evaporation is
unclear, but if it is naked, $R_e$ should be zero. Here it is assumed that
no remnant remains. For the massive black holes in the 
current epoch of the universe $R_e/H \ll 1$ will hold, with $H$ the Hubble
constant. The formation, complete evaporation or merging of black holes
thus resets the right-hand side of Equation (7) at some time $k_1$ and the 
matter-free evolution resumes until the next event.

Mass accretion and partial evaporation of black holes does not increase
the multiplicity of topological fluctuations associated with the handle
manifold. It does however increase or decrease the amplitude of the
fluctuation and thereby the effective scale on which matter and topology can
interact.

Equation (9) modifies the values of the prime multiplicities at some time
$k_1$, and subsequently the evolutionary path through state space.
Its solution derives from the motion of matter (see Equation (10) below) in
analogy with the coupled system formed by matter and metric in general
relativity: Mach's principle as interpreted by Einstein. Now one has,
{\it changes in topology and the local quantum motion of matter are determined
by the shape and global topology of space-time, and vice versa}.

\section{The Feynman Path Integral on a Multiply-Connected Space}

\subsection{The Quantum Foam}

The motion of matter (a world line) derives from a superposition of wave
amplitudes which are determined by the actions associated with the followed
paths. Imagine then an S$^3$ like space with matter
degrees of freedom. Suppose this space, with the quantum fields defined on it,
is contorted topologically into something which resembles an ensemble
of three-tori. The three-tori now add three more homotopic classes of paths.
Furthermore, the fields in quantum mechanics reflect the concept of
probability amplitudes spread over space. Rather than merely identifying the
edges of a region to create a closed three-torus topology, the extended nature
of the wave amplitudes should be preserved and the various three-tory connected
through three-ball surgery, i.e.~a connected sum. This results in four
homotopically inequivalent paths. In effect, a quadruplet of quantum fields
is constructed which will be called ``multiplication''.

As one performs the path integral in the quantum foam, the various classes with
fields are
indistinguishable, beyond their non-homotopicity, from one Planckian volume
to the other. As such, it should not matter in which order the different
group elements, i.e.~non-homotopic loops, are chosen at each ``point'' along
a path. This condition can only be satisfied if the group is perfect,
i.e.~$T^3$. Therefore, the construction above is the only one possible for a
single-type loop algebra.
{\it It is thus found that at the Planck scale the vacuum has
$e_{\rm M}=4$ topological degrees of freedom corresponding to a lattice of
three-tori}.

\subsection{The Propagator}

The inclusion of the quantum foam topology should yield a modified Feynman
propagator which has the correct asymptotic form on scales much larger than
$\ell_{\rm Planck}$. The following Ansatz is adopted for the amplitude
between space-time points $a$ and $b$
$$G(a,b)=\kappa \sum_j x_j \int_{\rm paths}
e^{iS_j^{ab_p}[m_j](x^{\mu})} Dx^{\mu}, \eqno (10)$$
where the sum over $j$ includes all topologically non-trivial primes and
$x_j$ is the multiplicity of prime ${\cal P}_j$ per Planckian volume.
Furthermore, $m_j$ denotes the number of homotopically inequivalent paths
between $a$ and $b$ for a given prime ${\cal P}_j$
$$m_j {\cal P}_j\equiv (T^{\dagger}T+1){\cal P}_j. \eqno(11)$$
This includes the path resulting from three-ball surgery.
Only the invariant ratios of the multiplicities of the non-trivial
primes $S^1\times S^2$ and $T^3$ enter in the amplitude of Equation (10).
The value of $\kappa$ is determined from
$$G(a,b)=\sum_c G(a,c)G(c,b). \eqno(12)$$

Given the number of homotopically inequivalent paths associated with a prime,
it is further assumed here that the wormhole associated with an $S^1\times S^2$
handle is not traversible by matter. That is, a particle path through the
wormhole has a divergent action if the particle's de Broglie
wavelength is smaller then the Schwarzschild radius. This issue of wormhole
traversibility is not settled, but it appears that the divergent Riemann
curvature associated with a black hole should be sufficient to completely
destroy the identity of the traversing matter.

\subsubsection{Particles}

The action in Planck units for a free relativistic particle of rest mass $m$
is proportional to the arclength
$$S^{ab_p}=-\hbar^{-1}m\int_p (g_{\mu\nu} dx^{\mu} dx^{\nu})^{1/2}, \eqno(13)$$
where all the symbols heve their usual meaning.
Even when no interactions between the fields occur on the intersections
where the three-tori are connected, the various paths of the homotopic classes
will merge. At these points, it appears that the physical system for which
the lagrangian needs to be written down consists of $e_{\rm M}$ identical
particles of mass $m$. Therefore, the action will increase by a factor
of $e_{\rm M}=m_j$, which leads to $S_j^{ab_p}[m_j]=m_j S^{ab_p}$
The number $m_j$ is then a multiplier in the exponent for the wave amplitude.
This is a classical rewording of the multiplication phenomenon and is useful
for semi-classical approximations.

\subsubsection{Fields}

More importantly, in a quantum field theory one wants to integrate over all
fields between two fixed field configurations on space-like surfaces and
include an interaction potential in the Lagrangian density. In this case the
action in Planck units for a single n-vector multiplet of scalar fields
$\phi_i$ and potential $V(\phi )$ would be of the form
$$S[\phi ]=\hbar^{-1}\int d^4x {{1}\over{2}}\partial_{\mu}\phi\partial^{\mu}
\phi-V(\phi ), \eqno(14)$$
where all the symbols have their usual meaning. For the three-torus lattice
one has $e_{\rm M}=4$ topological degrees of freedom. Hence,
{\it the interaction
potential is a fourth order polynomial of the scalar fields}
$$V(\phi )=\mu^2\phi_i\phi_i + \lambda (\phi_i\phi_i )^2, \eqno(15)$$
where $\lambda$ and $\mu^2$ are real. It is well-known
that these types of potentials are invariant under the action of the general
rotation groups O(n) and the unitary groups SU(n). Obviously, {\it these
symmetries are local and generate gauge theories}, although it is not apparent
that the $\phi_i$ are fundamental enough objects for a GUT (see also \S 5).

\subsection{Thermal Equilibrium of the Initial State}

The multiplicated matter degrees of freedom, can be absorbed by
black holes.
It is thus found that in the Planckian universe and in the presence of
sinks, i.e.~non-irreducible primes $S^1\times S^2$, a fraction of the
irreducible fluctuation energy is continuously being stored in mini black
holes. Simultaneuous black hole evaporation will then lead to local
thermalization at roughly the Planck temperature, as has been suggested
in[10].

Furthermore, for a large number of fluctuations and a $T^3$ lattice,
it follows from the central
limit theorem that the deviations from the Planck temperature are driven
to values much smaller than $M_{\rm Planck}$ under the action of
multiplication. The distribution function of the average sum of $n$
independent identically distributed Gaussian random variables of dispersion
$s=1$ in Planck units, has a total dispersion of $s/n^{1/2}$. The probability
for an $x=2s$ fluctuation is then approximately
$$p=e^{1/2}_{\rm M}e^{-1/2e_{\rm M}x^2s^{-2}}/(2\pi s^2)^{1/2}\approx 3\times 10^{-4}, \eqno(16)$$
more than two orders of magnitude smaller than in a simply-connected
space-time.

This process, being a direct consequence of the quantum foam, suppresses
fluctuations on scales larger than the particle horizon and is scale free by
nature. This explains the success of the Harrison-Zeldovich power spectrum
in reproducing the large scale structure of the universe.

\subsection{The Equivalence Principle}

The motion of matter as determined by Equation (10) depends on both general
relativistic and quantum interference effects.
The accelarations associated with the shape of the world line are now in part
caused by the non-trivial topological space-time. Still, the different
classes of paths can only be distinguished homotopically.
Therefore, the Gedanken experiment usually referred to as
``Einstein's lift'' still holds in small, but larger than $\ell_{\rm Planck}$,
regions of space. It is proposed then that the equivalence of inertial and
gravitational mass holds irrespective of the underlying $T^3$ topology.

Furthermore, no local measurement of the gravitational field can
distinguish one homotopic equivalence class from the other and Einstein
gravity remains valid. In a stronger sense, no local measurement of any
sort can make this distinction and the strong Principle of Equivalence should
hold. One may wonder how the required GL(4) symmetry of general relativity
can be facilitated. {\it The $T^3$ primes have an intrinsic size of
$\ell_{\rm Planck}$, so general relativity in the large scale and long time
limit can be envisaged as the way the lattice of three-tori bends and twists}.

\subsection{Evaluation of the Propagator}

In the (3+1) Arnowitt-Deser-Misner formulation of geometric dynamics, one
uses lapse and shift functions to integrate forward in time. The
Einstein equation is independent of topology and the action, say the arclength
or the integrated lagrangian density, in Equation (10) is covariant. The (3+1)
procedure then assures that no matter what lapse and shift functions are
chosen, the four-geometry, action, and the topology are evolved consistently
from one space-like slice to the next. When calculating the prime multiplicity
(an invariant) per Planckian volume, one should use the integral of the
covariant three-volume element $(det(g_{ij}))^{1/2}dx^i$. With the
three-metric $g_{ij}$ from the (3+1) decomposition of $g_{\mu\nu}$.

In practice, the path $a-b$ above should be much longer than the de Broglie
wavelength of some particle. So for de Broglie wavelengths larger than
the scale of a topological fluctuation, the propagator in Equation (10) will
limit to its classical value associated with the three-sphere.
For scales larger than $\ell_{\rm Planck}$ the three-tori are compactified
and only traversible wormholes could contribute to the interference effects.
If wormholes are traversible then no time difference is allowed between
their ``throats'' because as three-space is evolved, paths from future
directions will contribute to the wave amplitudes which is an obvious
violation of causality.
Of course, non-traversibility does not change the homotopical nature
of the wormholes.

\subsection{Correpondence Principle}

It is desirable that the Poincar\' e conjecture is true. Namely, that a
compact, simply-connected three-manifold is homeomorphic to the three-sphere.
If a fake three-cell were to exist, the structure of the algebraic 
equations or the form of the propagator (the motion of matter)
would not change. Also, the large scale limit provided by the Einstein equation
does not depend on the existence or non-existence of a fake three-cell.
Still, the theory would yield two or more non-homeomorphic[13], but loop
homotopic, classical limits. One would then need to speculate about the
physics, not represented by the loop algebra, to which this phenomenon
corresponds. Conversely, the fact that non-homeomorphic, but still homotopic,
primes exist in four dimensions and higher
(e.g.~$S^4\leftrightarrow S^2\times S^2$), suggests a topological origin for
the spatial dimension of the universe.

\subsection{Topological Amplification and GUT Inflation}

The standard model has a singularity which is conventially taken at time
$t=0$. As one approaches the sigularity, the temperature and density
diverge. Thus, no initial value problem can be defined at the singularity.
To avoid unknown quantum gravitational effects, the hot big bang scenario is
defined at a temperature comfortably below the Planck mass. This construction
requires extreme fine-tuning of the density and assumes thermal equilibrium
across $\sim 10^{80}$ causally disconnected regions.

Inflation theory provides a natural explanation for this flatness and horizon
problem[14]. Still, one is left with the question how to get to a universe in
approximate thermal equilibrium at a temperature of $10^{19}$ GeV with
sufficient energy stored in various degrees of freedom which are subsequently
inflated to yield a post-GUT universe. The multiplication mechanism with black
hole sinks of \S 3.2 provides a natural explanation for this initial condition.

It is clear from Equations (7) that the multiplicity of the fluctuations
$\alpha_k/\beta_k$ and $\gamma_k/\beta_k$ grows exponentially with time in the
absence of matter degrees of freedom. For the three-tori
this implies that even when the universe expands exponentially with an
e-folding time of $2t_{\rm Planck}$, their multiplicity per Planckian volume
will be of the order of unity for all time. Nevertheless, the solution is
dynamical and one might speculate about the effects of time reversal,
where the lattice is broken up in smaller isolated regions[15].

As the universe expands and cools away from its multiplicated initial
state, the multiplicity of the handle manifold will increase because
black holes cannot evaporate efficiently at very high initial temperatures
and primordial black holes will form through singular collapse.
Since the universe is not expanding exponentially with time,
merging events between black holes in the early universe will 
suppress the handle multiplicity. As the temperature drops, evaporation
becomes more and more efficient. In fact, for the choice $x=0$, the
formation of primordial black holes will coincide with the latter epoch[15].

The GUT symmetry should be broken when the average energy density of the
universe has dropped below threshold. Under the assumption that the exponential
increase in the scale factor of the universe is a robust consequence of
inflation, the handle multiplicity will become frozen in and the multiplicity
per Planckian volume will decrease to a value much smaller than unity.
Therefore, {\it at late times the value of $\alpha_n/\beta_n$ is determined by
the formation of macroscopic black holes}. This kind of result follows from
the fact that black holes can exist on all scales and therefore ``defy''
compactification when allowed to grow in mass through accretion and merging.

\section{Toy Models}

\subsection{Handles and Quantum Field Theory}

The initial values ${\cal X}=(x,y)=(0,1)$ yield a $1$-level
solution ${\cal T}=(\alpha_1/\beta_1,\gamma_1/\beta_1)=(1,7)$ if a
primordial black hole is formed.
Let $v\ell^3_{\rm Planck}$ denote the volume of the universe.
In the semi-classical approximation the propagator is proportional to
$$G\sim {{7}\over{v}}e^{4iS^c_1}+{{1}\over{v}}e^{iS^c_2}. \eqno(17)$$
For non-traversible wormholes $S^c_2\rightarrow\infty$, the additional
interference term in the propagator then has a rapidly varying
phase which prevents constructive interference. This correponds to an
effective optical depth through the quantum foam. 
This optical depth is roughly given by the integral of
${{\alpha_n}\over{\beta_n}}/v$ multiplied by the
effective radius of a black hole.
For values of the optical depth close to unity, thermalization as described
above will be effective.

In a full quantum field theory the following phenomenology then emerges.
In analogy with the above model, handles can connect onto
the $T^3$ lattice and generate additional interaction terms for the
fundamental symmetry group on the three-torus. These correspond to the extended
form of Mach's principle in which the global topology causes local topology
change. On the Planck scale, wormhole traversibility seems less of a burden on
causality and interference effects between the two primes can occur.

The meaning of the solution (7) is now also apparent. The $S^1\times S^2$
multiplicity per Planckian volume corresponds to the weights of the above
diagrams at various times.
From the discussion above, $\alpha_n/\beta_n<<1$ and handles should
provide a rather small term to the $T^3$ quantum field theory.
The lattice of three-tori in turn provides a number of symmetry groups for the
interacting fields. The SL(2;C), O(n) and SU(n) groups have already been
identified and \S 5 discusses the deeper ties with superstring theory.
In a companion paper[15], a more general model will be constructed which goes
beyond phenomenology.

\subsection{Black Hole Properties}

As a star collapses to a black hole, notwithstanding the possibility of a
stable equation of state for nuclear matter above neutronium densities,
it must reach a point where quantum effects become important. Topological
multiplication then distributes the vacuum over various degrees of freedom.
Although a black hole
has no (or very little) hair, one would expect this process to have a clear
macroscopic characteristic. It has been established above that $T^3$
is the only nuclear prime which is also a perfect group. Since in the late
time limit three-tori dominate the quantum foam, one has $e_{\rm M}=4$.

A black hole is a thermodynamic object which, from Bekenstein's
analogy, emits black body radiation of a temperature
$$T_{\rm BH}={{\kappa}\over{8\pi c}},\eqno(18)$$
where $\kappa=1/4M$ denotes the surface gravity and the constant $c$ is still
to be determined. One also has that the black hole entropy is given by
$$S_{\rm BH}=cA,\eqno(19)$$
with $A$ the surface area of the black hole which is uniquely determined
by its mass, charge and angular momentum. In fact, up to a numerical factor,
the latter quantities comprise all the information about the black hole.
That is, the entropy represents the total amount of information about the black
hole interior not accessible to observers. From the equivalence of the $T^3$
group elements beyond non-homotopicity it then follows that $c=1/e_{\rm M}$.

This result is the same as Hawking's derivation of the black hole temperature
[16]. The latter author did not make any assumptions about the quantum foam
topology, but showed that pair production could sustain an energy flow out
to infinity corresponding to a non-zero effective temperature of the black
hole. Only in a relativistic field theory can pair production occur. For this,
one has to require that the dominating Planck scale prime is also the boundary
of a Lorentz four-manifold SO(3,1;R), i.e.~is nuclear. Conversely, the perfect
group property alone makes $T^3$ unique, which derives purely from loop
homotopy, i.e.~superposition. Once more one is confronted with the fact that
black holes can exist on all scales and thereby form an inescapable link
between macroscopic and microscopic physics.

\subsection{Observational Diagnostics of the Quantum Foam}

As mentioned above, the current epoch is characterized by a $T^3$ lattice.
Furthermore, It is a prediction of the theory that the handle multiplicity at
late times is determined by the number of macroscopic black holes.

One can estimate that the current total number of black holes in the universe
is less than $10^{16}$. But if primordial black holes with masses of
roughly $10^{15}$ g were produced cupiously in the early universe, then this
number may be more than $p=20$ orders of magnitude larger and still be
consistent with observational constraints[17].

In any case, at the current epoch where the GUT fields have undergone their
phase transitions, the number of mini black holes spontaneously
created in the quantum foam divided by the total number of
Planckian volumes is a direct measure of the cosmological constant.
$\Lambda<10^{-168+p}$ $m_{\rm Planck}^2$, consistent with
the observational limit of $10^{-120}$ $m_{\rm Planck}^2$
without any fine-tuning.
Observations of both the Hubble constant
$H={\dot R}/R$ and the decelaration parameter $q_0=-{\ddot R}/RH^2$, with $R$
the scale factor of the universe, yield the cosmological constant. Since
an upper bound of $q_0<5$ yields $\Lambda/m_{\rm Planck}<10^{-120}$, accurate
knowledge of supernova rates as a function of time and the nature of the
remnant, neutron star or black hole, is extremely valuable.

One interesting thought is the following. The formation of
mini black holes is assumed to occur homogeneously over space.
Within this assumption, a number of events will occur within the Schwarzschild
radii of black holes and contribute to their mass $M$.
The magnitude of this effect should be proportional to $M^3$. Therefore,
the Schwarzschild radius of
an isolated, non-accreting, black hole of $10^9$ solar masses, should grow
by 1 part in $10^{19}$ over a period of one year, or by 1 part in $10^3$ over
a period of one year if the number of primordial black holes corresponds to
$p=16$. That is, cold black holes get colder.

\section{Discussion and Outlook}

In this work an algebraic equation has been derived and a basis set of prime
manifolds obeying a loop algebra constructed, in order to determine the
topological dynamics of the universe.
Einstein gravity has been extended into the quantum domain assuming that
the equivalence principle holds independent of topology. It has been shown
that the vacuum possesses four topological degrees of freedom corresponding
to a $T^3$ lattice. A physical boundary condition for the universe was found.
Natural solutions have been obtained for the thermal equilibrium of the initial
state, O(n) and SU(n) gauge groups for the fields, black hole entropy, and the
cosmological constant.

A problem to pursue further is the shape and amplitude
of the frozen-in quantum fluctuations after the epoch of inflation. These
may have observable consequences for the large scale structure of the universe
at various cosmological redshifts. Given the large and growing data sets
available on the history of structure formation in the universe, tight
constraints can be derived[15].

There appear to be a number of deep topological and group theoretical
connections. Adopting the superposition
principle in a topological manifold leads one to the loop homotopy which
yields a unique Planck scale structure through the three-torus. The $T^3$
topology assures Lorentz covariance or more general SL(2;C) invariance on the
bounded four-manifold. The latter group is the complexification of SL(2;R),
also known as the projection group. Recall that $T^3$ is a $Z_2$
extension of the discrete modular group SL(2;Z). Modular invariance in string
theory is crucial for supersymmetry and the finiteness of the multi-loop
amplitude. In fact, modular invariance forces self-duality of the string
lattice which restricts one to the $E_8\times E_8$ and Spin(32)/$Z_2$ groups
necessary for the anomaly cancellation in the 26-dimensional heterotic string.
These relations deserve further scrutiny.

A posteriori it appears that despite the obvious simplicity of the solutions
(7), they have a profound meaning. The very fact that the expressions only
involve three-tori and handles gives rise to a natural frame work in which to
do Planck scale quantum field theory. The solutions quoted above are
an indication of this. In \S 4.1 a phenomenological model was presented to
develop a full quantum field theory with a special role for the handle
manifold. The link with superstring theory suggested above seems very
interesting.

Obviously, many questions remain to be answered. The present work has been
concerned with the construction of an exploratory model to facilitate further
research, and a more complete theory will be presented in a companion paper.

The author is indebted to J.~Berendse-Vogels, W.~Berendse, E.~van 't Hof,
and M.~Bremer for stimulating discussions, and is grateful to the anonymous
referee for his comments which greatly helped to improve the presentation.
The author acknowledges with gratitude the support of NASA grant NAGW-3147
from the Long Term Space Astrophysics Research Program.

\end{document}